\documentclass[aip,reprint]{revtex4-1}

\setcitestyle{super}

\usepackage{graphicx}
\usepackage[caption=false]{subfig}
\usepackage{float}

\begin{document}

\title{Thickness and growth-condition dependence of \emph{in-situ} mobility and carrier density of epitaxial thin-film Bi$_2$Se$_3$} 

\author{Jack Hellerstedt}
\affiliation{School of Physics, Monash University, Victoria 3800, Australia}
\affiliation{Center for Nanophysics and Advanced Materials, University of Maryland, College Park, MD 20742-4111 USA}

\author{Mark Edmonds}
\affiliation{School of Physics, Monash University, Victoria 3800, Australia}

\author{J. H. Chen}
\altaffiliation{Present address: International Center for Quantum Materials, School of Physics, Peking University, Beijing 100871, P. R. China, and Collaborative Innovation Center of Quantum Matter, Beijing 100871, P. R. China}
\affiliation{Center for Nanophysics and Advanced Materials, University of Maryland, College Park, MD 20742-4111 USA}

\author{William G. Cullen}
\affiliation{Center for Nanophysics and Advanced Materials, University of Maryland, College Park, MD 20742-4111 USA}

\author{C. X. Zheng}
\affiliation{School of Physics, Monash University, Victoria 3800, Australia}

\author{Michael S. Fuhrer}
\email{michael.fuhrer@monash.edu}
\affiliation{School of Physics, Monash University, Victoria 3800, Australia}
\affiliation{Center for Nanophysics and Advanced Materials, University of Maryland, College Park, MD 20742-4111 USA}

\date{\today}

\begin{abstract}
Bismuth selenide Bi$_2$Se$_3$ was grown by molecular beam epitaxy while carrier density and mobility were measured directly \emph{in situ} as a function of film thickness.  Carrier density shows high interface \emph{n}-doping (1.5 x 10$^{13}$ cm$^{-2}$) at the onset of film conduction, and bulk dopant density of $\sim$5 x 10$^{18}$ cm$^{-3}$, roughly independent of growth temperature profile. Mobility depends more strongly on the growth temperature and is related to the crystalline quality of the samples quantified by \emph{ex-situ} AFM measurements.  These results indicate that Bi$_2$Se$_3$ as prepared by widely employed parameters is \emph{n}-doped before exposure to atmosphere, the doping is largely interfacial in origin, and dopants are not the limiting disorder in present Bi$_2$Se$_3$ films.
\end{abstract}


\maketitle

The topological insulator bismuth selenide (Bi$_2$Se$_3$) has a bulk band gap of 300 meV and single set of topologically non-trivial surface states, properties making it a good candidate to measure and exploit these states experimentally.\cite{Zhang2009}$^,$ \cite{Xia2009} Thin films of Bi$_2$Se$_3$ have been successfully grown by the van der Waals epitaxy method, \cite{Zhang2009a}$^,$ \cite{Li2010e}$^,$ \cite{Bansal2011b}$^,$ \cite{Taskin2012} where the crystal axis of the film aligns with that of the substrate, but the two lattices are incommensurate. However, even meticulously prepared bulk single crystals are persistently \emph{n}-doped,\cite{Butch2010} requiring substantial modification \cite{Kim2012} to bring the Fermi level into the bulk gap. Furthermore, exposure to atmosphere, specifically oxygen and water, has been demonstrated to increase the \emph{n}-type doping of the Bi$_2$Se$_3$ surface.\cite{Brahlek2011} However to date no electronic transport measurements have been reported on films that have never been exposed to ambient, thus leaving ambiguous whether observed doping was incurred during film growth or upon ambient exposure.

To address these issues of \emph{n}-doping and atmospheric exposure, we have developed an apparatus to grow thin films of Bi$_2$Se$_3$ by van der Waals epitaxy, while simultaneously allowing for the direct transport measurement of longitudinal and Hall sheet resistivities $\rho _{xx}$ and $\rho _{xy}$, from which we infer charge carrier sheet density $n$ = $B/e\rho _{xy }$ and mobility $\mu$ = 1/$ne\rho _{xx}$, where $B$ is the magnetic field and $e$ is the elementary charge.\cite{Hellerstedt2013}  Here we use this apparatus to examine the real time evolution of carrier density and mobility as a function of film thickness for different substrate temperature profiles. The dependence of transport parameters on film thickness have been reported previously for Bi$_2$Se$_3$ on Si(111) and Al$_2$O$_3$ substrates, \cite{Bansal2011}$^,$\cite{Kim2011}  however in those studies multiple films of difference thicknesses were measured \emph{ex situ}, and only one growth parameter regimen could be explored.

We study films grown using a two-temperature method \cite{Bansal2011b}$^,$ \cite{Taskin2012}$^,$ \cite{Zhang2011b} where the first two nanometers are grown at 110$^\circ$C and the remainder of the film grown at different temperatures between 200-230$^\circ$C.  We also study a single-temperature growth at 210$^\circ$C for comparison.  We find that the initial carrier density at the onset of conduction is approximately 1.5 x 10$^{13}$ cm$^{-2}$, and the carrier density growth with thickness corresponds to a bulk dopant density of $\sim$5 x 10$^{18}$ cm$^{-3}$, with both values roughly independent of growth conditions.  We conclude that a significant component of the observed doping in Bi$_2$Se$_3$ films is associated with the formation of the Bi$_2$Se$_3$-substrate and Bi$_2$Se$_3$-vacuum interfaces, even without exposure to ambient. In contrast, we find that the mobility depends strongly on growth conditions, with maximum mobility achieved for a two-step growth at 110$^\circ$C and 210$^\circ$C, consistent with the optimum growth temperatures observed by reflection high energy electron diffraction (RHEED) aided growth studies. \cite{Li2010e}$^,$\cite{Bansal2011b}$^,$ \cite{Taskin2012}$^,$\cite{Bansal2011}$^,$\cite{Kim2011}$^,$\cite{Zhang2011b} The results are strongly suggestive that the charge carrier mobility in our Bi$_2$Se$_3$ films (and likely in the films prepared in other laboratories) is not limited by ionized impurities (dopants) but rather by structural uniformity. Atomic force microscopy (AFM) confirms a strong correlation between mobility and structural coherence of the films.

\begin{figure}[!]
\centering
\begin{tabular}{c}
	\subfloat{\includegraphics[keepaspectratio=true,width=1\linewidth]{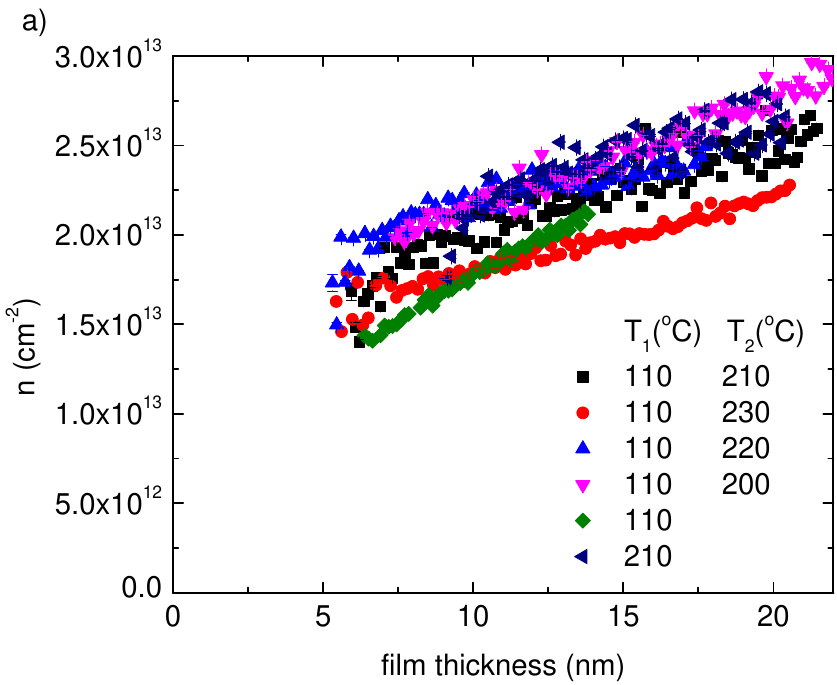}} \\
	\subfloat{\includegraphics[keepaspectratio=true,width=1\linewidth]{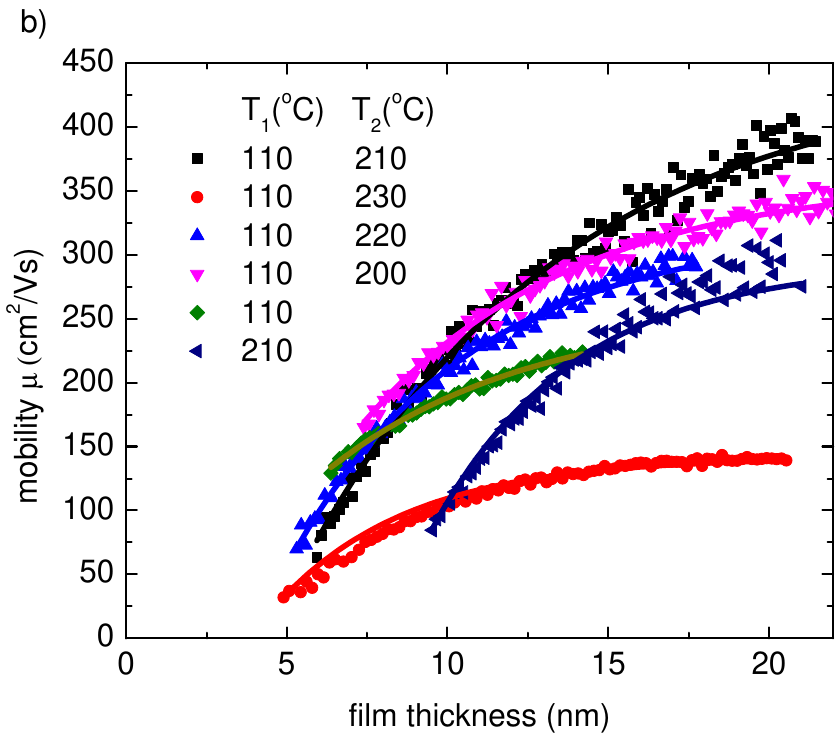}}
\end{tabular}

\caption{(a) Hall carrier density $n$ and (b) mobility $\mu$ versus Bi$_2$Se$_3$ film thickness for various growth temperature profiles as shown in legend: \emph{T}$_1$ is the substrate temperature for the first 2nm of film deposited, \emph{T}$_2$ is the temperature for the remainder of the growth. In the cases where there is no \emph{T}$_2$ the \emph{T}$_1$ temperature was maintained for the whole growth.}
\label{n-mobility-thickness-fig}
\end{figure}

SrTiO$_3$ [111] (STO) substrates (Shinkosha, Princeton Scientific) are prepared by flow-through oxygen annealing at 1050$^\circ$C for 4-6 hours to achieve an atomically flat surface.\cite{Chang2008a}  The details of our growth technique and \emph{in-situ} electrical measurements have been previously described in greater detail.\cite{Hellerstedt2013} Briefly, bismuth and selenium are co-deposited from Knudsen cells (at rates of 0.1 nm/min and 1.8 nm/min, respectively) through a stencil mask in a Hall bar configuration onto pre-patterned Ti (5 nm)/Au (50 nm) electrical contacts.  Two-stage growth is carried out at a temperature of 110$^\circ$C for the first two nanometers of film deposition, before increasing the substrate temperature to a series of `hot' temperatures between 200 and 230$^\circ$C. Two additional samples were grown: in one instance the substrate temperature was maintained at 110$^\circ$C, for the other, the substrate temperature was 210$^\circ$C the entire time, forgoing the initial low temperature growth.  Longitudinal and Hall resistivities are monitored continuously during film growth by periodic application of a magnetic field swept continuously between $\pm$ 1100 Gauss.  The thickness of each sample is calculated from the bismuth rate using the empirically determined tooling factor for the growth chamber.

\begin{figure}[!]
\centering
\includegraphics[keepaspectratio=true,width=1\linewidth]{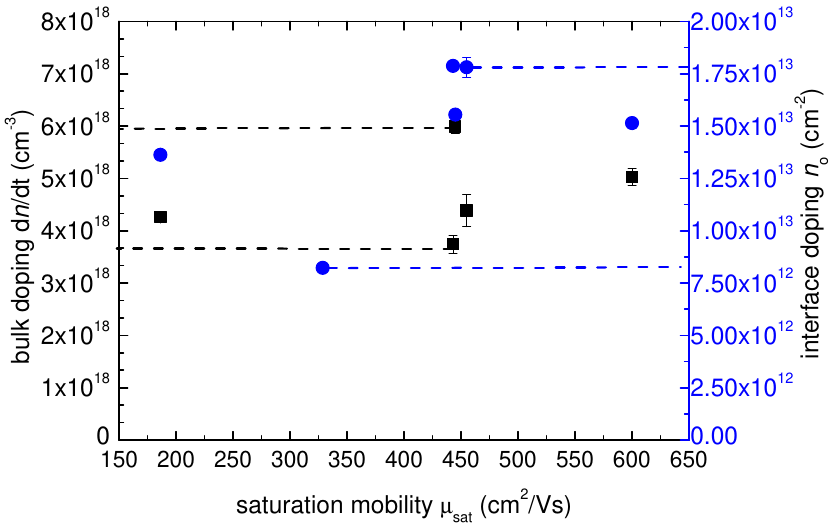}
\caption{Bulk dopant concentration d$n$/d$t$ (squares, left) and interfacial dopant concentration $n_0$ (circles, right) extracted from the slope and intercept, respectively, of linear fits to $n(t)$ in fig. \ref{n-mobility-thickness-fig}(a), plotted against saturation mobilities extracted from the data in fig. \ref{n-mobility-thickness-fig}(b).}
\label{bulk-interface-n-fig}
\end{figure}

Figure \ref{n-mobility-thickness-fig}(a) shows $n(t)$ and fig. \ref{n-mobility-thickness-fig}(b) shows $\mu(t)$ for the different growth temperature profiles (we observe large fluctuations in the magnitude of $\rho_{xy}$ with time during the initial growth when $\rho _{xx}$ $>$ 100 k$\Omega$, typically for the first 5 nm of film growth, where we presume the film is highly non-uniform in thickness or possibly discontinuous. We therefore report data for which $\rho_{xx}$ $<$ 100 k$\Omega$, for which $\rho_{xy}$ values become stable and we presume the film is continuous and uniform). As shown in fig. \ref{n-mobility-thickness-fig}(a), $n(t)$ is very similar for all growth temperature profiles studied. $n(t)$ is roughly linear, with a positive intercept at zero thickness. The results are in contrast to \emph{ex-situ} measurements of Bi$_2$Se$_3$ on Si(111) where $n \sim \sqrt{t}$,\cite{Kim2011} and Bi$_2$Se$_3$ on Al$_2$O$_3$ where $n$ is independent of thickness.\cite{Bansal2011} We performed a linear fit to the data in fig. \ref{n-mobility-thickness-fig}(a) and interpret the intercept $n_0$ $\approx$ (1-2) x 10$^{13}$ cm$^{-2}$ as an interface dopant concentration which is present as soon as the film is formed, and the slope d$n$/d$t$ $\approx$ 5 x 10$^{18}$ cm$^{-3}$ as a bulk (3D) dopant concentration which scales linearly with thickness indicating a constant rate of new dopants added as new film is deposited. We cannot deduce from our measurements whether the interface dopants correspond to the substrate-Bi$_2$Se$_3$ or Bi$_2$Se$_3$-vacuum interface, however the \emph{in-situ} nature of our measurements shows that the doping is inherent to the interfaces and not due to reaction with ambient.

Figure \ref{n-mobility-thickness-fig}(b) shows the thickness dependence of the mobility $\mu(t)$.  In contrast to $n(t)$, $\mu(t)$ shows appreciable changes with different growth temperature profiles.  Note that the reported mobilities are measured at the respective growth temperatures, explaining their somewhat lower magnitude relative to mobilities reported elsewhere in the literature. Solid lines in fig. \ref{n-mobility-thickness-fig}(b) are fits to the measurement using the phenomenological relation \cite{Kim2011}:

\begin{equation}
\mu(t)=\frac{\mu_{sat}}{1+\frac{\lambda^*}{t-t_0}}
\label{mu-thickness-equation}
\end{equation}

where $\mu_{sat}$ is interpreted as the saturation mobility in the thick-film limit. We find that $\mu_{sat}$ varies by a factor of 3 for different growth conditions, while the carrier concentration at a given thickness varies by less than a factor of 1.5. This result is surprising, since the carrier concentration is expected to reflect the concentration of ionized impurities (dopants), which have been suspected as a major source of disorder (i.e. lowered mobility) in Bi$_2$Se$_3$.\cite{Kim2012} Instead, we find that mobility varies strongly with growth conditions though dopant concentrations are nearly constant.

To further explore the correlation between dopant concentration and mobility, fig. \ref{bulk-interface-n-fig} shows the bulk dopant concentration d$n$/d$t$ and interfacial dopant concentration $n_0$ extracted from fig. \ref{n-mobility-thickness-fig}(a) as a function of the $\mu_{sat}$ extracted from fig. \ref{n-mobility-thickness-fig}(b). There is no clear correlation of either d$n$/d$t$ or $n_0$ with $\mu_{sat}$, in fact the highest and lowest-mobility films have very similar concentrations of both bulk and interfacial dopants. We conclude that dopants are not the dominant disorder in our films, and that the changes in mobility with growth conditions are controlled by another variable.

\begin{figure}[!]
\centering
\begin{tabular}{c}
	\subfloat{\includegraphics[keepaspectratio=true,width=1\linewidth]{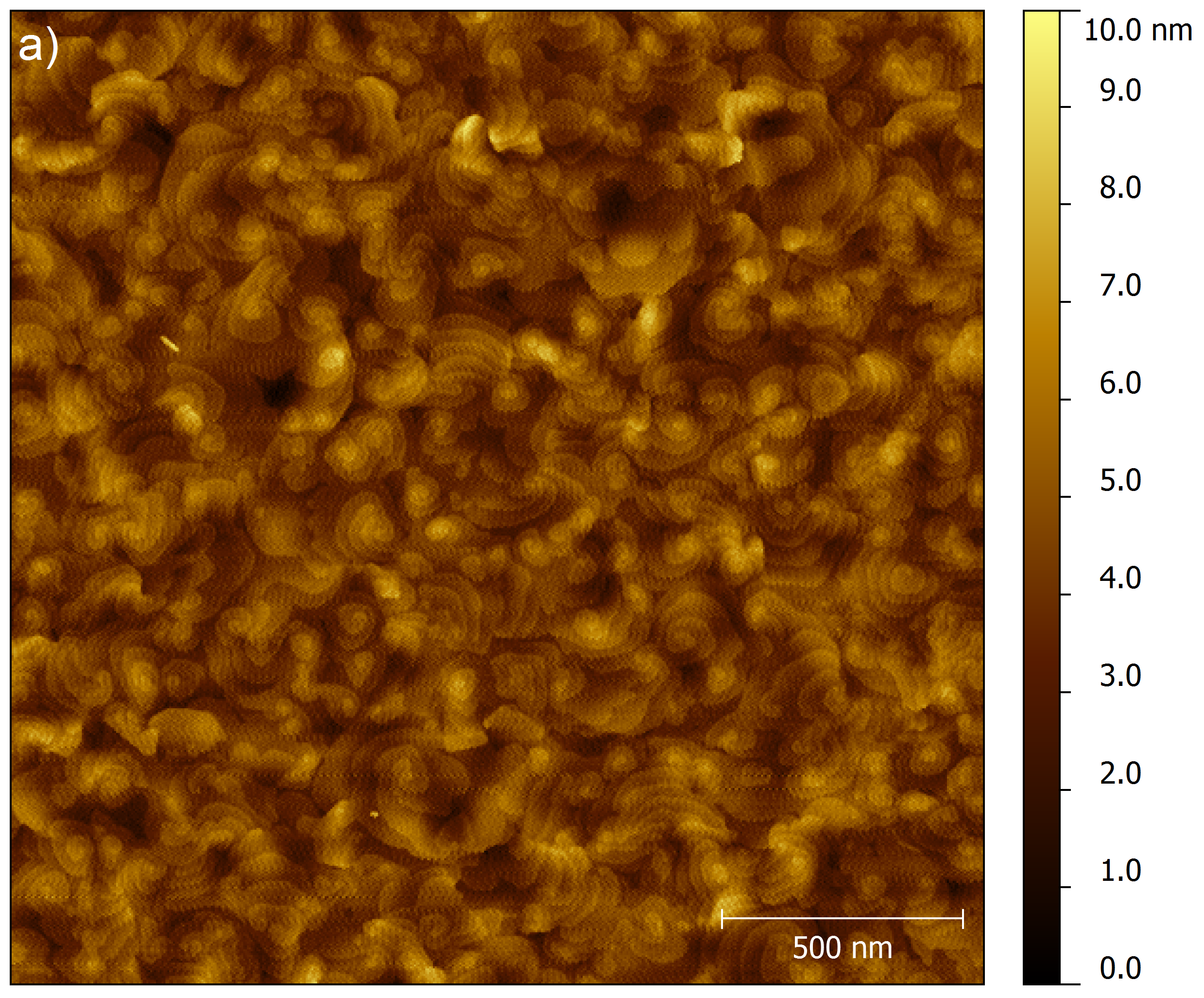}} \\
	\subfloat{\includegraphics[keepaspectratio=true,width=1\linewidth]{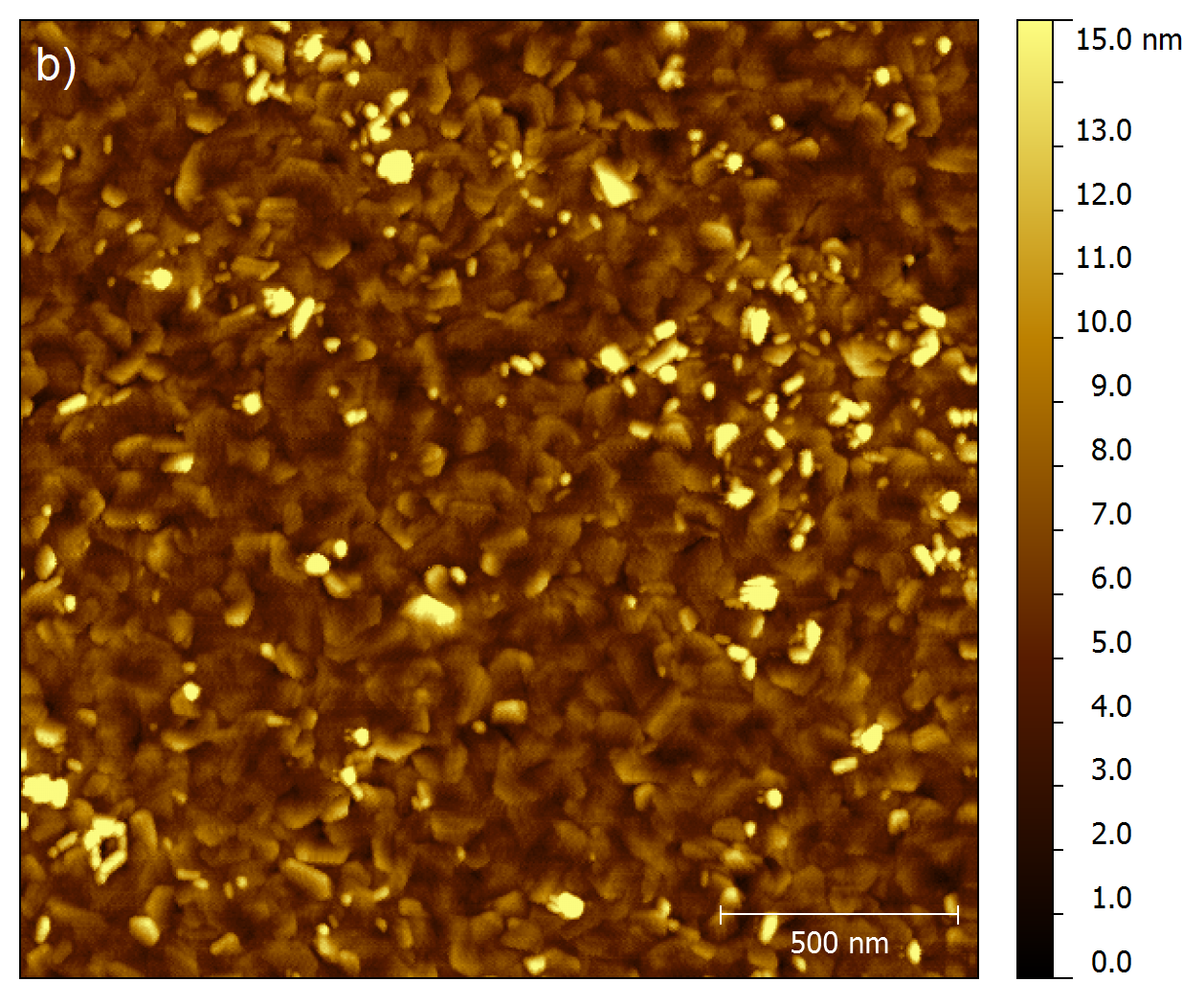}}
\end{tabular}

\caption{\emph{Ex-situ} 2$\mu$m x 2$\mu$m atomic force microscope images of the a) 110-210$^\circ$C and b) 110-230$^\circ$C films.}
\label{AFM-fig}
\end{figure}

\begin{figure}[!]
\centering
\begin{tabular}{c}
	\subfloat{\includegraphics[keepaspectratio=true,width=1\linewidth]{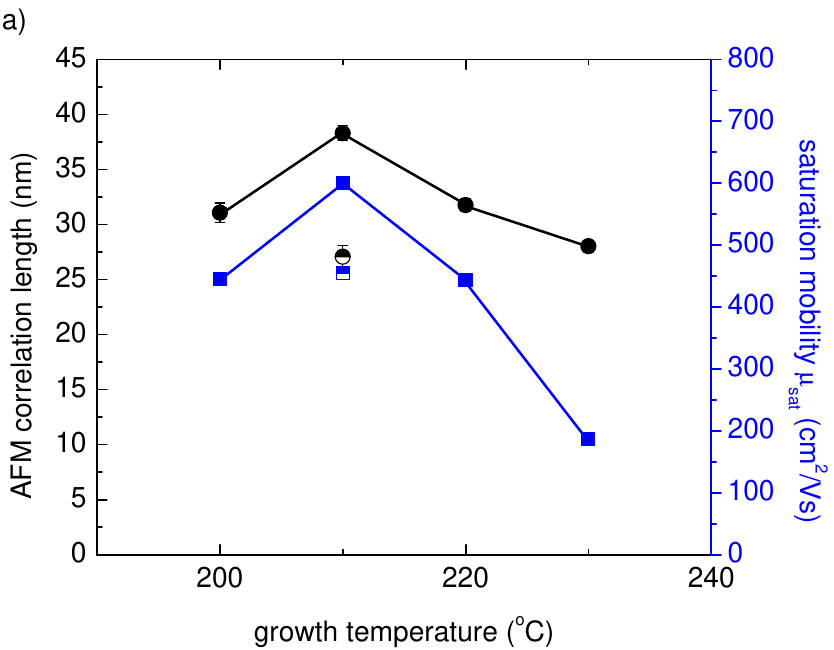}} \\
	\subfloat{\includegraphics[keepaspectratio=true,width=1\linewidth]{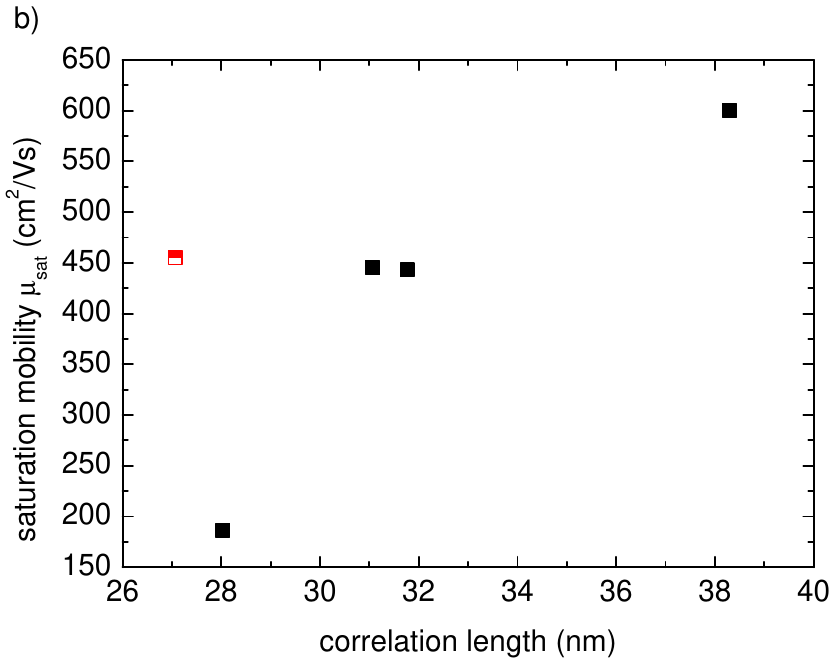}}
\end{tabular}

\caption{(a) Saturation mobilities $\mu_{sat}$ (squares) extracted from the data in fig. \ref{n-mobility-thickness-fig}(b) and correlation lengths $T$ (circles) from AFM images are plotted as a function of the highest temperature during growth for each respective sample. The half filled points at 210$^\circ$C correspond to the single-step growth sample. (b) Saturation mobility plotted against AFM correlation length for each sample; black points are two-step growths, the red half-filled point is the single-step growth at 210$^\circ$C.}
\label{mobility-correlation-temp-fig}
\end{figure}

To this end, we chose to further investigate the crystal quality of the films as another possible variable controlling mobility.  Figure \ref{AFM-fig} shows representative \emph{ex situ} atomic force microscopy (AFM) images taken using a Bruker Dimension Icon using the contact mode of imaging. A clear difference is immediately evident between the two images: the sample grown at 110-210$^\circ$C for which mobility is optimal shows evidence of van der Waals epitaxial growth, with atomically flat terraces of width $\sim$50 nm, while the sample grown at 110-230$^\circ$C with lower mobility shows greater roughness and evidence of some crystallites with c-axes oriented non-parallel to the film normal.  The differences are quantified using the height-height correlation function $H(x)$ (function of point-to-point distance $x$) computed for each sample and fit to:

\begin{equation}
f(x)=2 \sigma^2(1-e^{-(x/T)^2})
\label{HHCF-equation}
\end{equation}

where $\sigma$ is the standard deviation in height, and $T$ is the surface correlation length. We assume $T$ is a measure of the structural coherence of the film. For an epitaxial film, $T$ would be proportional to the mean terrace width.

Figure \ref{mobility-correlation-temp-fig}(a) shows the AFM correlation length and saturation mobility $\mu_{sat}$ as a function of the hot growth temperature for each sample, showing a peak both in the mobility and the correlation length for the two-step growth at 110-210$^\circ$C. The single-step 210$^\circ$C growth values are also shown, measurably lower than the two-step growth terminating at the same temperature. To further explore this relationship, in fig. \ref{mobility-correlation-temp-fig}(b) we show $\mu_{sat}$ plotted against the AFM correlation length $T$. A high degree of correlation between $\mu_{sat}$ and $T$ is observed for the two-step samples, strongly indicating that the structural coherence of the film is the driving variable behind mobility variation with growth conditions.

In this work we have demonstrated that bismuth selenide films grown according to widely employed methodology within the literature are substantially \emph{n}-doped as grown in vacuum, before any exposure to atmosphere. Significant interface ($\sim$(1-2) x 10$^{13}$ cm$^{-2}$) and bulk ($\sim$ 5 x 10$^{18}$ cm$^{-3}$) dopant concentrations are observed, nearly independent of growth conditions.  Differences in mobility with respect to growth temperature appear to be driven by changes in film morphology, as quantified by \emph{ex-situ} AFM analysis, rather than variations in dopant concentrations. The implication is that work to date on Bi$_2$Se$_3$ film preparation has optimized the structural coherence of the films, but significant gains remain to be made in reducing both bulk and interfacial dopant concentrations.

\begin{acknowledgments}
This work was supported in part by the NSF grants DMR-11-05224 and DMR-05-20471 (Maryland MRSEC), and performed in part at the Melbourne Centre for Nanofabrication (MCN) in the Victorian Node of the Australian National Fabrication Facility (ANFF). MSF is supported by an ARC Laureate Fellowship.
\end{acknowledgments}

\bibliography{manuscript} 

\end{document}